# Iran's scientific dominance and the emergence of South-East Asian countries as scientific collaborators in the Persian Gulf Region

Henk F. Moed

Independent researcher and senior scientific advisor, Amsterdam, The Netherlands. Email: hf.moed@gmail.com

## Abstract

A longitudinal bibliometric analysis of publications indexed in Thomson Reuters' Incites and Elsevier's Scopus, and published from Persian Gulf States and neighbouring Middle East countries, shows clear effects of major political events during the past 35 years. Predictions made in 2006 by the US diplomat Richard N. Haass on political changes in the Middle East have come true in the Gulf States' national scientific research systems, to the extent that Iran has become in 2015 by far the leading country in the Persian Gulf, and South-East Asian countries including China, Malaysia and South Korea have become major scientific collaborators, displacing the USA and other large Western countries. But collaborations patterns among Persian Gulf States show no apparent relationship with differences in Islam denominations.

## 1. Introduction

Political developments in the Persian Gulf Region are still in the centre of global public interest. A commentary published in Nature in 1991 shortly after the start of Operation Desert Storm analysed international scientific collaboration patterns during the 1980s of the 8 Persian Gulf States (Iran, Iraq, Kuwait, Saudi Arabia, Qatar, Bahrain, United Arab Emirates and Oman) and 4 neighbouring Middle East countries (Jordan, Lebanon, Syria and Yemen), and compared these patterns with changes in international political relations (De Bruin, Braam and Moed, 1991). Its base assumptions stated that international scientific collaboration patterns reflect geographical, political, social and historical relations (Frame and Carpenter, 1979) and that it is important for all stakeholders to have a thorough understanding of the relationships in an area of political tension.

This short communication provides an update of the 1991 study by De Bruin, Braam and Moed. It presents a longitudinal bibliometric analysis of publications published from the 8 Persian Gulf States and 4 neighbouring Middle East countries, and indexed in Thomson Reuters' Web of Science and Elsevier's Scopus, covering a time period as long as 35 years (1980-2015). The US diplomat Richard N. Haass predicted in 2006 trends in what he termed as the upcoming "Middle East fifth era" (Haass, 2006). This paper empirically examines four of Haass' key predictions, namely that (i) "the United States will continue to enjoy more influence in the region than any other outside power, but its influence will be reduced from what it once was"; (ii) "United States will increasingly be challenged by the foreign policies of other outsiders"; (iii) "Iran will be one of the two most powerful states in the region"; and (iv) "tensions between Sunnis and Shiites will grow throughout the Middle East".

## 2. A bibliometric model of scientific development

Science policy analysts need appropriate tools to monitor the state of their country's scientific development, enabling them to roughly categorize countries in terms of scientific development and compare "like with like" rather than directly compare them on the basis of absolute numbers, regardless of their phase of development. For analysts from scientifically developing countries such numbers are often of limited value and can even be discouraging as they are extracted from global rankings which tend to position scientifically developed countries rank in a ranking's upper part. They need an assessment framework based on the notion that the scientific and economic conditions in which a lesser-developed country finds itself at a certain moment are not static but can be viewed as a phase in a process that more scientifically developed countries have already entered with great success.

Table 1. A bibliometric model for capturing the state of a country's scientific development

| Phase | Description | Trend in # published articles | Trend in % internationally co-authored publications |
|-------|-------------|-------------------------------|----------------------------------------------------|
| Pre-development | Low research activity without clear policy of structural funding of research | ~ | ~ |
| Building-up | Collaborations with developed countries are established; national researchers enter international scientific networks | + | ++ |
| Consolidation and expansion | The country develops its own infrastructure; the amount of funds available for research increases | ++ | - |
| Internationalization | Research institutions in the country start as fully-fledged partners, increasingly take the lead in international collaboration | + | + |

Legend to Table 1. ~ denotes: no clear trend; +: increase; - : decline; ++: strong increase. Source: UNESCO (2014). For more information on this model, see Moed and Halevi (2014).

A bibliometric model of scientific development meeting this requirement is presented in Table 1. It is applied in the analysis presented in the next sections. It distinguishes four different phases of development of a national research system: (i) a pre-development phase; (ii) building up; (iii) consolidation and expansion; and (iv) internationalization. These phases are briefly described in Table 1. The model assumes that during the various phases of a country's scientific development, the number of published articles in peer-reviewed journals shows a more or less continuous increase, although the rate of increase may vary substantially over the years. The share of a country's internationally co-authored articles, however, discriminates between the various phases in the development. Table 1 reflects the discontinuities that the model assumes that take place in the indicator values over time when moving from one phase into another. The model is experimental and needs to be further validated and expanded with citation impact indicators, which can be used in the fourth phase of the development process to monitor the internationalization process.

## 3. Trends in the number of publications and internationally co-authored publications

Figure 1 presents the annual number of publications during 1980-2014 of the 8 Persian Gulf States, 4 neighbouring Middle East countries (Jordan, Lebanon, Syria and Yemen), and for three benchmark countries: Egypt, Turkey and Israel. It shows that Qatar and United Arab Emirates (UAE) had the largest increase in publication counts during 1980-2014: three orders of magnitude, and Kuwait and Iraq the smallest (zero and one order of magnitude, respectively). The latter outcome suggests that Kuwait has never overcome the devastations of the 1990-1991 Persian Gulf War. Iraq recuperated to some extent after the 2003 invasion by a United States-led coalition (Operation Iraqi Freedom) and the deposition of the Ba'athist government of Saddam Hussein, as it started revealing moderate positive annual growth rates as from 2005.

Iran and Saudi Arabia had in 2014 the largest absolute number of publications of all Persian Gulf States, namely 29,000 and 13,000, respectively. Iran's research output declined during the first half of the 1980s under the influence of the Iraq-Iran War which started in September 1980, but as from the beginning of the 1990s, when the Persian Gulf War started with Iraq's invasion of Kuwait, it revealed an exponential growth, doubling up until 2011 approximately every 3 years. Saudi Arabia's annual publication counts showed almost flat growth rates during 1990-2007, but rapidly increased to around 50 per cent in 2010-2011, but then declined to 20 per cent in 2013-2014.

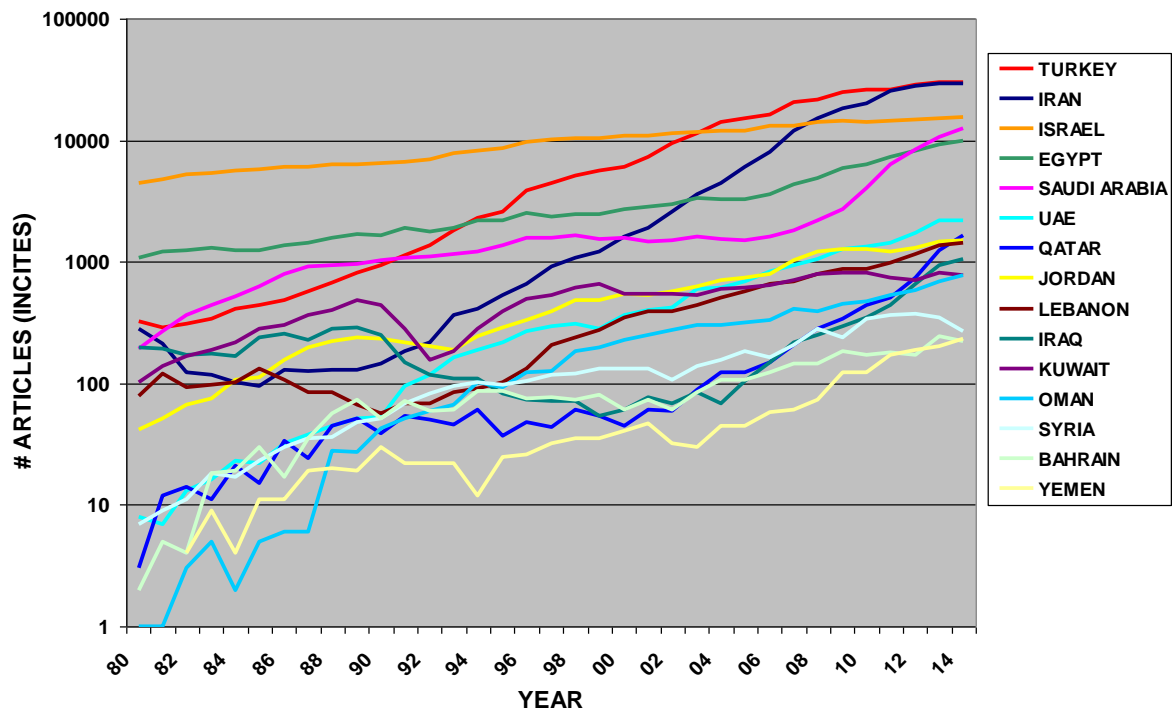

Figure 1: Annual number of publications during 1980-2014 for the 8 Persian Gulf States and 4 neighbouring Middle East countries and 3 benchmark countries. Data were extracted from Thomson Reuters' Incites. Since data for 2015 were not yet complete at the date of data collection, they are not displayed. The number of 2015-articles indexed up until 5 December 2015 amounts to 20,400 for Iran against 19,600 for Turkey.

According to data extracted from Scopus, using its subject classification into 27 disciplines, for USA, UK and other larger Western countries medicine tends to be the most important discipline, with typically 20 percent of publications, followed by engineering and biochemistry, genetics & molecular biology, each with some 10 per cent. But in China and Malaysia these percentages are reversed, while India, Pakistan and South Korea have an intermediary position. The latter is also true for most Persian Gulf States. The share of their papers in medicine and in biochemistry, genetics & molecular biology is in most cases lower than that of larger western countries, and that in engineering, physics & astronomy and chemistry higher. Exceptions are Bahrain and Lebanon, with 28 and 24 percent of articles in medicine, respectively.

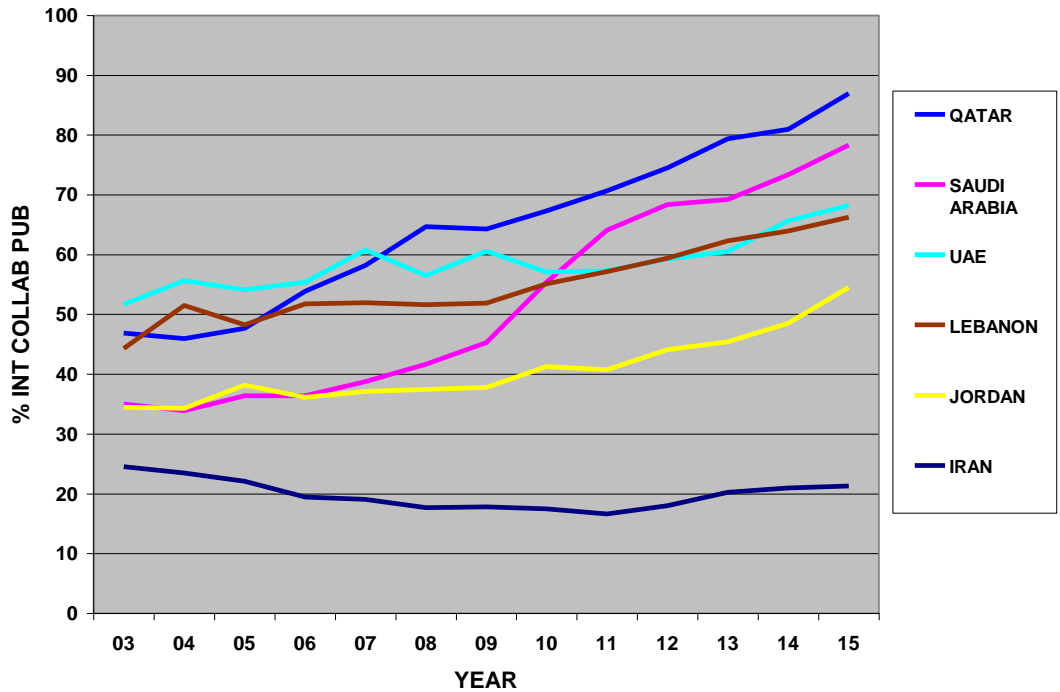

Figure 2. Percentage of internationally co-authored publications (relative to total publication output) of the 6 most frequently publishing Persian Gulf States or neighbouring Middle East countries. Data from Elsevier's Scopus.

Large differences exist between Iran and Saudi Arabia with respect to the amount of foreign input needed to produce these papers. According to Figure 2, the percentage of internationally co-authored publications (ICAP) is in 2015 almost 80 per cent for Saudi Arabia, but only around 20 per cent for Iran. Applying the model of scientific development presented in Section 2, the results suggest that these two countries are in different phases of scientific development. While Saudi Arabia and most other Persian Gulf States are still in the building-up phase, Iran is currently moving from a consolidation and expansion into the internationalization phase. This trend can be expected to continue now that the international boycotts towards this country are cancelled. The other Persian Gulf States still depend in various degrees upon collaboration with external institutions, increase their ICAP rate and are, in terms of the scientific development model presented in Table 1, in a phase of building up a scientific infrastructure. This is especially true for the two countries showing the largest increase of their publication output, namely Qatar and UAE, with ICAP percentages in 2015 of 87 and 68, respectively.

Gringas (2014) found that a disproportionally large number of researchers appearing in the Thomson Reuters' list of highly cited researchers indicate Saudi universities as secondary address, thus boosting these institutions up in global university rankings. The articles in which this occurs are counted as internationally co-authored publications in Figure 2 and can be expected to boost up the Saudi percentage of ICAP as well. Even if the influence of this phenomenon is substantial, it underlines the dependence of this country upon the input of foreign researchers and does therefore not violate the conclusions on its state of scientific development.

Haass' third prediction states that "Iran will be one of the two most powerful states in the region" (Haass, 2006). During the past 3 decades, the only country that has been able to create and expand a research infrastructure of its own is Iran, despite the economic boycotts to which it has been subjected during most of the time. In terms of scientific development, Iran is clearly the leading country in the Gulf region. As from 2007, Iran's annual publication count in Incites exceeded both that of Egypt and Israel, and in 2015, – based on an analysis of about 50 per cent of the total number of 2015-articles eventually published –, also Turkey.

More bibliometric indicators for a large set of Middle East and neighbouring countries for the total time period 1981-2013 and derived from Incites can be found in Gul et al. (2015). Although this study does not analyze trends over the years, it does present citation impact indicators, showing that Israel ranks first with an impact relative to the world of 1.13 while for all other countries it is below 0.6. This outcome suggests that although Israel and Iran are both in the internationalization phase of the development process, in terms of generating citation impact the first country is currently still far ahead of the second.

## 4. International scientific collaboration patters

Figures 3 and 4 present a detailed analysis of the international scientific collaborations involving the same 12 countries (8 Persian Gulf States and 4 neighbouring Middle East countries) as those analysed in De Bruin, Braam and Moed (1991). Figure 3 presents a VOS map of scientific collaborations among the countries in this set of 12 nations. Similar to Multi-Dimensional Scaling (MDS), VOS aims to locate items in a low-dimensional space in such a way that the distance between any two items reflects the similarity of the items as accurately as possible, but differs from MDS in the way in which it attempts to achieve this aim (Van Eck, Waltman, Dekker & Van den Berg, 2010). The clustering model is a variant of modularity based clustering, a technique aiming to maximize a modularity measure of a network, defined as the fraction of the links that fall within a given group minus the expected such fraction if links were distributed at random. The VOS technique is a weighted and parameterized variant able to detect small clusters or communities (Waltman, Van Eck & Noyons, 2010). Located at the left hand side is a community with five countries with Shia dominance either within a country's Muslim population or in its government (Syria and Yemen). In the remaining two clusters all countries but one have a Sunni dominance. The striking exception is Iran in the right hand cluster, in which 90 per cent of population is Shia.

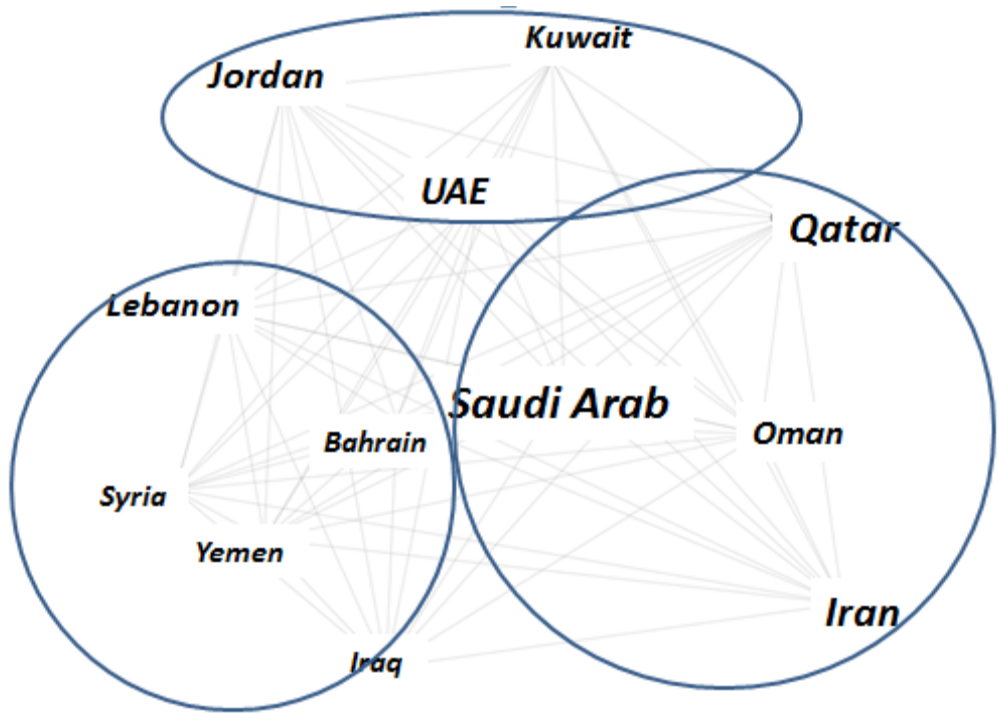

**Figure 3.** VOS Viewer map of the international co-authorship relations among 8 Persian Gulf States and 4 neighbouring Middle East countries for the year 2015. Data were extracted from Thomson Reuters' Incites. The three circles are inserted by the author of this paper and indicate clusters as identified in the VOS clustering module.

The community structure in the international co-authorship network among the 12 Gulf and neighbouring Middle East countries displayed in Figure 3 can be partially interpreted as traces of the main dominations within the Islam, with United Arab Emirates, Lebanon and Saudi Arabia as bridges between the two. But Iran's striking appearance in 2015 in a cluster of countries in which the majority of the population is Sunni ("Islam", n.d.) does not seem to align with Haass' assertion that "tensions between Sunnis and Shiites will grow throughout the Middle East". Intensifying collaborations with Sunny dominated countries does align with a strategy by Iran aiming to become a member of the Arab League, a regional organization of currently 22 Arab countries to which Iran has applied for membership ("Arab League", n.d.).

Analyzing collaborations between Persian Gulf States and 4 neighbouring Middle East countries on the one hand and countries outside the Middle East region on the other (Figure 4), the most striking feature is the emergence of East and South Asian collaboration partners during the past decade, namely China and South Korea in the East, Malaysia in the Southeast and Pakistan and India in the South. Malaysia has in 2015 strong links both with Iran and Iraq, while the other four South-East Asian countries have links with Saudi Arabia only. Iran shows a strong orientation towards Northern America and Western Europe; Malaysia is the only South-East Asian country with which Iran is linked in the map. Saudi Arabia shows a more balanced position towards Western and Asian countries as it has links with five South-East Asian countries, but also with 7 Western countries. The two emerging countries Qatar and UAE have the strongest ties with the USA.

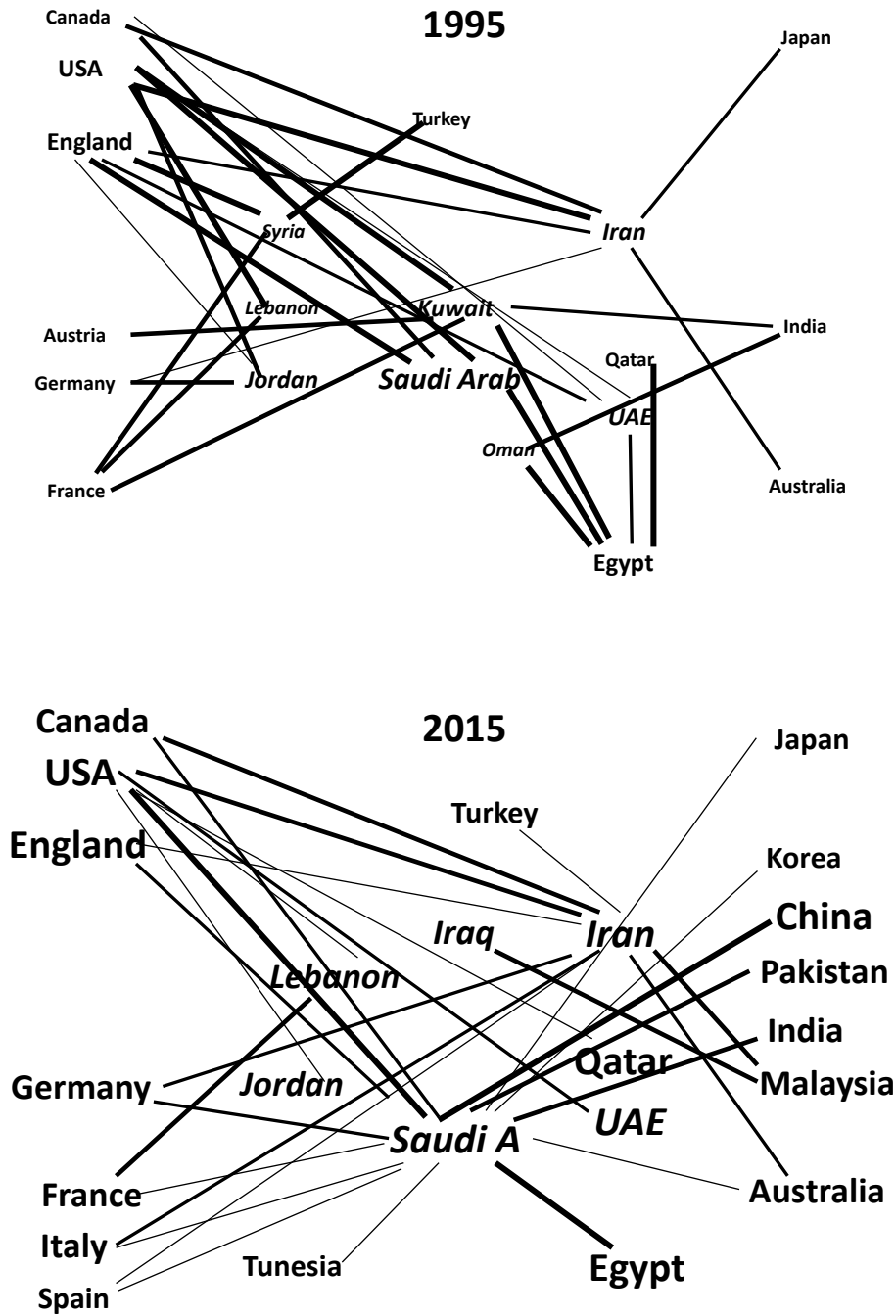

<u>Figure 4</u>. Collaboration ties (ICAP) between Persian Gulf and neighbouring Middle East States and countries outside the Middle East region for the years 1995 and 2015. Data are extracted from Incites. Figures are constructed in the same manner as those presented in De Bruin, Braam and Moed (1991) for the 1980s. They show the 30 strongest links in a particular year. Countries are positioned in a topological map. Font size indicates the number of co-authorship links, by grouping countries into quartiles on the basis of the number of co-publications in any year. The thickness of the lines the strength of the collaboration (Salton's index), applying a similar quartile approach.

Co-authorship links between Persian Gulf States and countries outside the region are typically one order of magnitude stronger than those among Gulf States. Russia does not appear in Figure 4 or Table 2. The overall level of international scientific collaboration in this country is still low, due to historical factors, although it is increasing (Kotsemir et al., 2015). In its collaborations with the Persian Gulf States and neighbouring Middle East countries, Russia switched from Syria in 1995 to Saudi Arabia and Iran in 2015.

Table2: Co-authorship strength between Gulf or neighbouring Middle East states and countries outside the Middle East

| Rank | Country | Mean co-authorship strength in 2015 | % Change compared to 2005 |
|------|---------|-------------------------------------|---------------------------|
| 1 | USA | 0.077 | -25 % |
| 2 | England | 0.052 | -34 % |
| 3 | Malaysia | 0.050 | +138 % |
| 4 | France | 0.048 | -26 % |
| 5 | Egypt | 0.048 | -19 % |
| 6 | Germany | 0.044 | -26 % |
| 7 | Canada | 0.043 | -31 % |
| 8 | China | 0.039 | +48 % |
| 9 | India | 0.039 | -2 % |
| 10 | Italy | 0.037 | +2 % |
| 11 | Australia | 0.036 | -2 % |
| 12 | Turkey | 0.033 | -8 % |
| 13 | Spain | 0.032 | +8 % |
| 14 | Pakistan | 0.031 | +32 % |
| 15 | South Korea | 0.030 | +106 % |

Legend to Table 2. Co-authorship strength between two countries is defined as the number of co-authorship links between them weighted on their total number of co-authorship links (Salton's Index). Underlying data were extracted from Thomson Reuters' Incites.

Table 2 shows that in 2015 the USA is still the most important external scientific partner in the Persian Gulf region and neighbouring Middle East countries. But compared to the situation in 2005, the strength of the average co-authorship relation between USA and the 12 Gulf and neighbouring Middle East States declined with 25 per cent. Four South-East Asian countries, Malaysia, China, Pakistan and South Korea show large positive growth rates. While before 2008 China's ties with Iran and Saudi Arabia were of similar strength, in 2011 China started showing a preference for the latter country; in 2015 the strength of their ties is more than twice that between China and Iran. The tie between China and Saudi Arabia is among the three strongest in the region in 2015. Only the ties of the latter country with Egypt and with USA are stronger. In 2009, China became the largest importer of oil from the Gulf, but also surpassed the United States as the largest single exporter to the region as well. China's main oil provider is actually Saudi Arabia (Wakefield & Levenstein, 2011).

Table 3. Preferred foreign collaborators of Gulf States by discipline in 2014.

| Gulf State | Collab country | discipline | Gulf State | Collab country | discipline |
|---|---|---|---|---|---|
| IRAN | Canada | Computer Science | SAUDI ARABIA | Canada | Computer Sci |
| | Germany | Physics & Astron | | | Medicine |
| | Italy | Physics & Astron | | China | Mathematics |
| | Malaysia | Environmental Sci | | France | Physics & Astron |
| | Spain | Agr & Biol Sci | | Germany | Agr & Biol Sci |
| | | Physics & Astron | | | Physics & Astron |
| | Turkey | Mathematics | | India | Pharmacol, Toxicol |
| | | Physics & Astron | | Italy | Physics & Astron |
| | UK | Physics & Astron | | Japan | Physics & Astron |
| JORDAN | USA | Medicine | | S Korea | Chemical Eng |
| LEBANON | France | Chemistry | | | Materials Science |
| | | Physics & Astron | | | Physics & Astron |
| | USA | Medicine | | Spain | Mathematics |
| QATAR | USA | Physics & Astron | | | Physics & Astron |
| | | | | Tunesia | Computer Sci |
| | | | | | Physics & Astron |

Legend to Table 3. Data were extracted from Scopus, using its subject classification into 27 disciplines, and relate to the year 2014.

Table 3 shows the external countries the Gulf and neighbouring Middle East countries preferred to collaborate with in the various disciplines in 2014. All collaboration pairs displayed in Figure 4 (year 2015) were analyzed. Per discipline a ratio was calculated of the percentage of internationally co-authored papers between a Gulf State and a foreign country and this gulf state's overall percentage of co-authorships with any country. Table 3 includes only cases for which this ratio exceeds 1.5.

The outcomes presented in Figure 4 and Table 2 fully align with Haass' first and second trend, namely that "the United States will continue to enjoy more influence in the region than any other outside power, but its influence will be reduced from what it once was", and that "United States will increasingly be challenged by the foreign policies of other outsiders". Other large Western countries show a decline as well. Instead, Malaysia, China, Pakistan and South Korea have substantially increased the collaboration in the Gulf during the past 10 years. Malaysia is now even the third country partner in the Gulf.

## 5. Concluding remark

International scientific collaboration patterns reflect geographical, political, social and historical relations and may also actively contribute to shaping these, and thus have an effect upon political relations as well. The recently established bilateral trade agreement of $600 billion between China and Iran in the next decade (Sharafedin, 2016) illustrates how Iran's dominance and the increasing role of main South-East Asian countries in the scientific development of the Gulf States have a clear correlate in the political domain.

## Acknowledgement

The author wishes to thank Thomson Reuters for providing access to Incites, and the Sapienza University of Rome for offering its library facilities, especially access to Elsevier's Scopus.com.

**References**

Arab League (n.d.). From https://en.wikipedia.org/wiki/Arab_League-Iran_relation.

De Bruin, R.E., Braam, R.R., Moed, H.F. (1991). Bibliometric lines in the sand. *Nature* **349**, 559-562.

Frame, J.D. & Carpenter, M. P. (1979). International research collaboration. Social Studies of Science, 9, 481-497.

Gringas, Y. (2014). How to boost your university up the rankings. University World News. 18 July 2014 Issue No:329.
http://www.universityworldnews.com/article.php?story=20140715142345754

Gul, S., Nisa, N., Shah, T.A., Gupta, S., Jan, A., Ahmad, S. (2015). Middle East: Research productivity and performance across nations. Scientometrics. 105:1157–1166.

Haass, R.N. (2006). The New Middle East. Foreign Affairs.
https://www.foreignaffairs.com/articles/middle-east/2006-11-01/new-middle-east.

Islam (n.d.). Information was obtained from a series of articles available at https://en.wikipedia.org/wiki/Islam_in_X, X being the name of the country.

Kotsemir M., Kuznetsova T., Nasybulina E., Pikalova A. (2015) Identifying Directions for the Russia's Science and Technology Cooperation. Foresight and STI Governance, 9, 54–72.

UNESCO (2014). Higher Education in Asia: Expanding Out, Expanding Up. ISBN 978-92-9189-147-4 licensed under CC-BY-SA 3.0 IGO. Montreal: UIS. http://www.uis.unesco.org.

Moed, H.F. and Halevi, G. (2014). Tracking scientific development and collaborations – The case of 25 Asian countries. Research Trends, Issue 38, September 2014.

Sharafedin, B. (2016). Iran's leader says never trusted the West, seeks closer ties with China. Markets, Sat 23 January, 2016.
http://www.reuters.com/article/us-iran-china-idUSKCN0V109V.

Van Eck, N.J., Waltman, L., Dekker, R., & Van den Berg, J. (2010). A comparison of two techniques for bibliometric mapping: Multidimensional scaling and VOS. Journal of the American Society for Information Science and Technology, 61(12), 2405-2416.

Wakefield, B., Levenstein, L. (eds.) ( 2011). China and the Persian Gulf: Implications for the United States. Woodrow Wilson International Center for Scholars, Washington, D.C..

Waltman, L., Van Eck, N.J., & Noyons, E.C.M. (2010). A unified approach to mapping and clustering of bibliometric networks. Journal of Informetrics, 4(4), 629-635.
.